\documentclass[aps,pre,twocolumn,floatfix,showpacs]{revtex4}
\usepackage{graphicx}

\begin{document}
\title{Wave-number Selection by Target Patterns and Side Walls in Rayleigh-B\'{e}nard Convection} 
\author{John R. Royer}
\author{Patrick O'Neill}
\author{Nathan Becker}
\author{Guenter Ahlers}
\affiliation{Department of Physics and iQUEST,\\ University of
California, Santa Barbara, CA  93106}
\date{ \today} 

\begin{abstract}
We present experimental results for patterns of Rayleigh-B\'{e}nard convection in a cylindrical container with static side-wall forcing. The fluid  used was methanol, with a Prandlt number $\sigma = 7.17$, and the aspect ratio was  $\Gamma \equiv R/d
 \simeq19 $ ($R$ is the radius and $d$ the thickness of the fluid layer). 

In the presence of a small heat input along the side wall, a sudden jump of the temperature difference $\Delta T$ from below to slightly above a critical value $\Delta T_c$ produced a stable pattern of concentric rolls (a target pattern) with the central roll (the umbilicus) at the center of the cell. 
A quasi-static increase of $\varepsilon \equiv \Delta T/\Delta T_c - 1$  beyond $\varepsilon_{1,c} \simeq 0.8$ caused the umbilicus of the pattern to move off center. As observed by others, a further quasi-static increase of $\varepsilon$ up to $\varepsilon = 15.6$ caused a sequence of transitions at $\varepsilon_{i,b}, i = 1, ..., 8$, each associated with the loss of one convection roll at the umbilicus. Each loss of a roll was preceded by the displacement of the umbilicus away from the center of the cell.  After each transition the umbilicus moved back toward but never quite reached the center. With decreasing $\varepsilon$ new rolls formed at the umbilicus when $\varepsilon$ was reduced below $\varepsilon_{i,a} < \varepsilon_{i,b}$. When decreasing $\varepsilon$, large umbilicus displacements did not occur.
 
In addition to quantitative measurements of the umbilicus displacement, we determined and analyzed  the entire wave-director field of each image.  The  wave numbers varied in the axial direction, with minima at the umbilicus and at the cell wall and a maximum at a radial position close to $2 \Gamma / 3$. The wave numbers at the maximum showed hysteretic jumps at $\varepsilon_{i,b}$ and $\varepsilon_{i,a}$, but on average agreed well with the theoretical predictions for the wave numbers selected in the far field of an infinitely extended target pattern. To our knowledge there is as yet no prediction for the wave number selected by the umbilicus itself, or by the cell wall of the finite experimental system.

\end{abstract}

\maketitle

\section{Introduction}
\label{sec:introduction}

When a temperature difference $\Delta T$ exceeding a critical value $\Delta T_c$ is  applied across a thin, horizontal layer of fluid,  convection occurs. \cite{Ch61} This phenomenon is know as Rayleigh-B\'{e}nard convection (RBC). 
 The fluid flow then  forms a pattern. 
The onset of convection is determined by the Rayleigh number
\begin{equation}
R = \alpha g d^3 \Delta T / \kappa \nu\ .
\end{equation}
Here $\alpha$ is the isobaric thermal expansion coefficient, $g$ the acceletation of gravity, $d$ the thickness of the sample, $\kappa$ the thermal diffusivity, and $\nu$ the kinematic viscosity. Linear stability-analysis for a laterally infinite system shows that convection will occur with a critical wave number $k_c = 2\pi/\lambda_c = 3.117$ ($\lambda$ is measured in units of $d$) when $R$ exceeds $R_c = 1707.3$.  \cite{Ch61} 
For the laterally infinite system the patterns that evolve beyond the onset of convection and their stability are determined by $R$ and by 
the Prandtl number
\begin{equation}
\sigma = \nu / \kappa\ .
\end{equation}
Weakly nonlinear theory predicts that immediately above onset,  a laterally infinite uniform system should yield a pattern consisting of parallel, straight rolls.\cite{SLB65} For real experimental systems the patterns are  influenced by the lateral boundaries which contain the fluid. Even in these finite systems excellent approximations  of the predicted ideal straight rolls of the infinite system can be found under some conditions. \cite{BPA00,FNrolls,BMCA99} However, the boundaries can also lead to patterns of different symmetry. Thus,
by heating a thin cylindrical convection cell gently from the side in addition to heating primarily from below,  a pattern of {\it concentric} rolls can be stabilized. A concentric pattern can also be the result of horizontal temperature gradients near a cylindrical side wall which are intrinsic to the particular construction of the sample cell.  The concentric pattern will consist of $n$ rolls, where $n$ is an integer. Of these, the one in the middle really would be more properly viewed as a convection {\it cell}, with either up-flow or down-flow in its center. It is often referred to as the umbilicus. This type of pattern, know also as a target pattern, was studied extensively by Koschmieder and Pallas \cite{KP74} for relatively large $\sigma$. The rolls have a mean dimensionless wave length $\bar \lambda = 2\Gamma / n$ where $\Gamma =  R/d$ ($R$ is the sample radius) is the aspect ratio of the sample and where $n$ is the number of rolls along a radius. The corresponding wave number is $\bar k \equiv 2 \pi / \bar \lambda =  \pi n / \Gamma$. Koschmieder and Pallas found that the wave length of the rolls away from the center and the side wall increased with increasing $\Delta T$. They also observed a sequence of transitions with increasing $\Delta T$, each of which involved the loss of a roll at the sample center. 

Later work for smaller $\sigma$ showed  that at a certain value $\varepsilon_{1,c}$ of $\varepsilon \equiv R/R_c - 1$ the target pattern becomes unstable. \cite{CP84,SAC85,CLPG86,CGP86,HEA93} The instability is known as the focus instability. \cite{NPS90} How this occurs in detail depends on $\sigma$, $\Gamma$,  and on the nature of the confining side walls. Here we describe what happens for our particular  sample ($\sigma = 7.17, \Gamma = 19$), where the side walls fixed the phase of the pattern at the radial position $r = \Gamma$ ($r$ is scaled by the cell thickness $d$). For the work of Ref.\cite{SAC85} ($\sigma = 6.1, \Gamma = 7.5$) the observed phenomena were similar. As $\varepsilon$ is increased in small steps from small values, the umbilicus shrinks. The result is a gradual decrease  of the local wave numbers of the rolls away from the center. When $\varepsilon$ is increased beyond $\varepsilon_{1,c}$, the umbilicus breaks the cylindrical symmetry by moving off center. This transition is continuous and the off-center patterns are stationary in time, but the radial umbilicus position is $\varepsilon$ dependent.  As $\varepsilon$ is further increased, the umbilicus becomes even smaller, and then disappears at $\varepsilon = \varepsilon_{1,b}$. At that point the pattern once more approaches a state of cylindrically symmetric  rolls, although perfect rotational invariance is not fully recovered. With further increase of $\varepsilon$, this process repeats itself, with another roll loss at $\varepsilon_{2,b} > \varepsilon_{1,b}$, and so forth. The net result is a gradual reduction with increasing $\varepsilon$ of the wave numbers of the rolls between transitions and a discontinuous change of $\bar k$  at each transition.   The transitions are hysteretic, occurring at $\varepsilon_{i,a} <  \varepsilon_{i,b}$ when $\Delta T$ is decreased. 

We studied target patterns for  $0 < \varepsilon < 15.6$ in a sample with $\Gamma = 19.0$ using methanol with $\sigma = 7.17$. We developed high-resolution numerical umbilicus-detection algorithms and determined quantitatively the umbilicus displacement $\delta$ with increasing as well as decreasing $\varepsilon$. At small $\varepsilon$ the results revealed the location of the focus instability, at $\varepsilon \simeq 0.8$. Previous measurements, \cite{HEA93} for $\sigma = 0.93$ and a larger $\Gamma = 43$, had found the instability at lower values, near $\varepsilon \simeq 0.1$. Over  the whole $\varepsilon$ range of our experiment the data for $\delta$, together with measurements of an average wave number $\langle k \rangle$ away from the center and the side wall,  revealed eight hysteretic transitions. We observed large umbilicus displacement only for increasing $\varepsilon$; for decreasing $\varepsilon$ the hysteresis yielded transition points $\varepsilon_{i,a}$ which were sufficiently low to avoid the $\varepsilon$-range over which the umbilicus displacement was large. 

For $\varepsilon > 5.6$ cross rolls  \cite{CB74,BC79} formed at the outermost roll when the system was close to a transition at $\varepsilon_{i,b}$. This was particularly pronounced on the side opposite to the umbilicus displacement where the local wave lengths were exceptionally large when $\delta$ was large. Nonetheless, the general nature of the pattern was maintained up to our largest $\varepsilon$ values. For $\varepsilon \agt 14$ periodic time dependence associated with the oscillatory instability \cite{CB74,BC79} developed in the region near the cross rolls. 

A complete characterization of the pattern involves a knowledge of the entire wave-director field $\vec k(r,\theta)$ where $\theta$ is the angular and $r$ the radial position in the sample. Rather than simply measuring an average wave number along a cell diameter (as was done in previous work), we implemented a local wave-director analysis \cite{HG87,CMT94,EMB98} and determined $\vec k(r,\theta)$. From  $\vec k(r,\theta)$ we could then determine various averaged quantities, including $\langle k \rangle$. 
 The results for $\langle k \rangle$, determined over a radius range that excluded the rolls near the center and the side wall, were discontinuous at the transitions. The hysteresis loop of each transition was traced out quantitatively. We found that the azimuthal average $k_\theta(r) \equiv \langle k \rangle _\theta$ had interesting structure as a function of $r$, showing different selection at the umbilicus, in the bulk of the sample, and at the cell wall. The competition between the selection mechanisms yielded a broad maximum of  $k_\theta(r)$ near $r =  2 \Gamma/3$. For our experimental conditions it did not yield any radially traveling waves, as predicted by Tuckerman and Barkley \cite{TB88,BT89} for the case of conducting side walls. As was found in previous work, \cite{CP84,SAC85,CLPG86,HEA93} the values of $\langle k \rangle$ selected in the bulk were in good agreement with the predicted wave numbers for the far field of infinitely extended concentric rolls  when the discontinuities at the transitions were smoothed out. We also report results for the wave numbers selected at the umbilicus and at the side wall; but for these there seem to be no theoretical prediction.

In the next section we shall review theoretical predictions for the selection by concentric rolls and relevant previous experiments. Then. in Sect.~\ref{sec:experiment}, we shall discuss the experimental apparatus and procedures, as well as the image-analysis methods. Section \ref{sec:results} gives our results. It consists of a discussion of the patterns observed in various $\varepsilon$ ranges, of a presentation of the umbilicus-displacements results, and of a presentation of our wave-number results. A brief section summarizing this work ends the paper.

\section{Theoretical predictions and previous Experiment}
\label{sec:theory}

\begin{figure}
 \includegraphics[width=8.5cm]{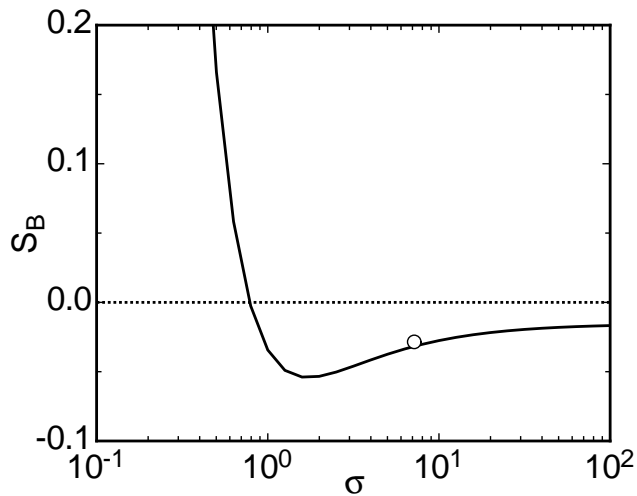}
\caption{The theoretical prediction for $S_B = (k_B/k_c - 1)/\varepsilon$ in the limit of small $\varepsilon$ as a function of the Prandtl number $\sigma$. The circle is the result of the resent work to be discussed below.}
\label{fig:S_B}
\end{figure}

Theoretical predictions of the wave numbers $k_B(\varepsilon)$ selected in the far field  of infinitely extended target patterns have been made by several authors. \cite{PM81,Cr83,MP83,BC86,NPS90} For the infinite system wave-number adjustment can take place by expansion of the pattern to large distances. Finite laboratory systems differ from this in an important way. The side wall, under typical experimental conditions, pins the phase of the pattern and prevents this expansion. Thus, a change of the average wave number with changing $\varepsilon$ is possible only at the umbilicus. However, also at the umbilicus unhindered phase slip is not possible. Instead, a  discontinuous and hysteretic process involving the destruction or creation of a convection cell occurs   and leads to a  discontinuous change of the wave-number field at the transition. The discontinuous effect on the average wave number decreases as the aspect ratio of the sample increases because the loss of a single cell in the center is a smaller perturbation for a larger sample. Thus, even for the sample with a boundary one expects, in the large-$\Gamma$ limit,  a continuous curve $k_B(\varepsilon)$ as a function of $\varepsilon$. Theoretically the location of this curve is determined by the rotational symmetry of the target pattern which does not permit the mean flow that is induced by roll curvature under less symmetric conditions. 
Thus, the mean flow  must  be balanced precisely by a pressure gradient.\cite{PM81,Cr83,MP83} This condition leads to a unique wave number $k_B(\varepsilon)$. Initial predictions of the selected wave numbers were applicable only for small $\varepsilon$ and are given by \cite{Cr83,MP83}
\begin{equation}
k_B/k_c - 1 = S_B \varepsilon + {\cal O}(\varepsilon^2)
\end{equation}
 with $S_B = -N'/R_2$, $N' = 0.1659 + 1.426/\sigma - 1.220/\sigma^2$, and $R_2 = 10.76 - 0.073/\sigma + 0.128 / \sigma^2\varepsilon$. Here $k_c =3.117$ is the critical wave number at the onset of convection. In Fig.~\ref{fig:S_B} we show $S_B$ as a function of $\sigma$.  For the present work we have $\sigma = 7.17$ and $S_B = -0.0317$. The measurements to be described below yielded the experimental value $-0.0285\pm 0.001$. This result is shown as the circle in the figure. It falls slightly above, but is generally in good agreement with, the prediction. For the work of Ref.~\cite{CP84} ($\sigma = 14$) the theory yields $S_B = -0.0243$, but no corresponding result has been extracted from the data. Measurements for $\sigma = 0.93$ were reported in Ref.~\cite{HEA93} and yielded a value for $S_B$ close to zero, but a precise number was not quoted. For that case one obtains the prediction $S_B = -0.0266$, but $S_B(\sigma)$ changes rapidly with $\sigma$ and passes through zero at $\sigma = 0.79$.

\begin{figure}
 \includegraphics[width=8.5cm]{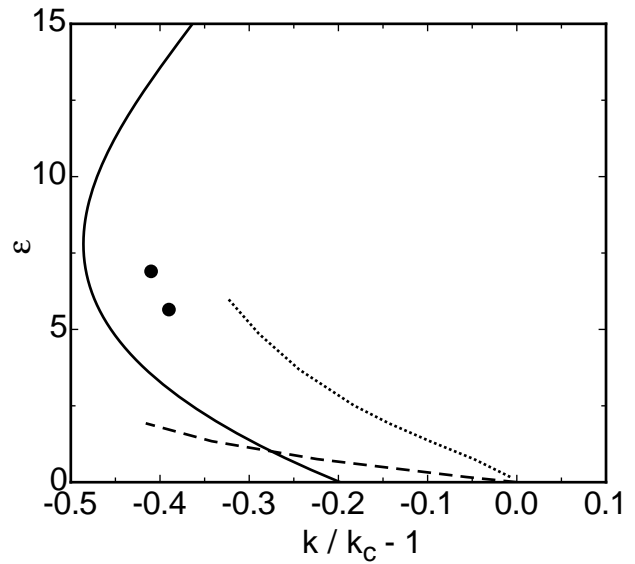}
\caption{Stability boundaries of infinitely extended straight rolls for a Prandtl number $\sigma = 7.0$. Dashed line: zigzag instability \protect \cite{BC79}. Solid line: cross-roll instability  \protect \cite{BC79}. Also shown, as a dotted line, are the wave numbers predicted for the far field of infinitely extended concentric rolls \protect \cite{BC86}. The two solid circles correspond to our experimental observation of the cross-roll instability near the side wall.}
\label{fig:bussball}
\end{figure}

Calculations valid at larger $\varepsilon$ were made by Buell and Catton \cite{BC86} (BC) and by Newell et al. \cite{NPS90} using a combination of analytic and numerical methods. BC gave results for a few values of $\sigma$ and $R$, and we shall compare our measurements with interpolations between these predictions. We show their results for a Prandtl number close to that of our experiment in Fig.~\ref{fig:bussball}. Also shown in that figure are the nearby stability boundaries of infinitely extended uniform straight rolls.\cite{CB74,BC79} Although in the large-$\sigma$ limit it is expected that the zig-zag instability will coincide with the selected wave number,\cite{PM81,Cr83,MP83} it is clear from the figure that $\sigma = 7$ is still far from that limit. 

To our knowledge there is as yet no prediction of the wave-number selection at the umbilicus itself. The results to be presented here show that the wave number selected there is smaller than the one in the far field.

Aside from the phase pinning mentioned above, the side walls of experimental cells have an additional influence on the pattern. The boundary conditions imposed by the walls can yield a separate and competing selection mechanism.\cite{CDHS,TB88,BT89} This mechanism is expected to depend on the conductivity of the side walls relative to that of the fluid, and it is not understood quantitatively for typical experimental conditions. For perfectly conducting side walls, Tuckermann  and Barkley \cite{TB88,BT89} predicted a pattern of radially-traveling waves. The traveling nature of the pattern can be understood in terms of  the competition between the selection by the curved rolls on the one hand and by the wall on the other which can lead to a wave-number gradient which in turn can lead to a non-zero time derivative of the phase of the pattern. \cite{FNTW2,RBWB87} However, to our knowledge such radially traveling waves have not yet been observed in experiments. \cite{FNTW,TBA02}

\begin{figure}
 \includegraphics[width=7cm]{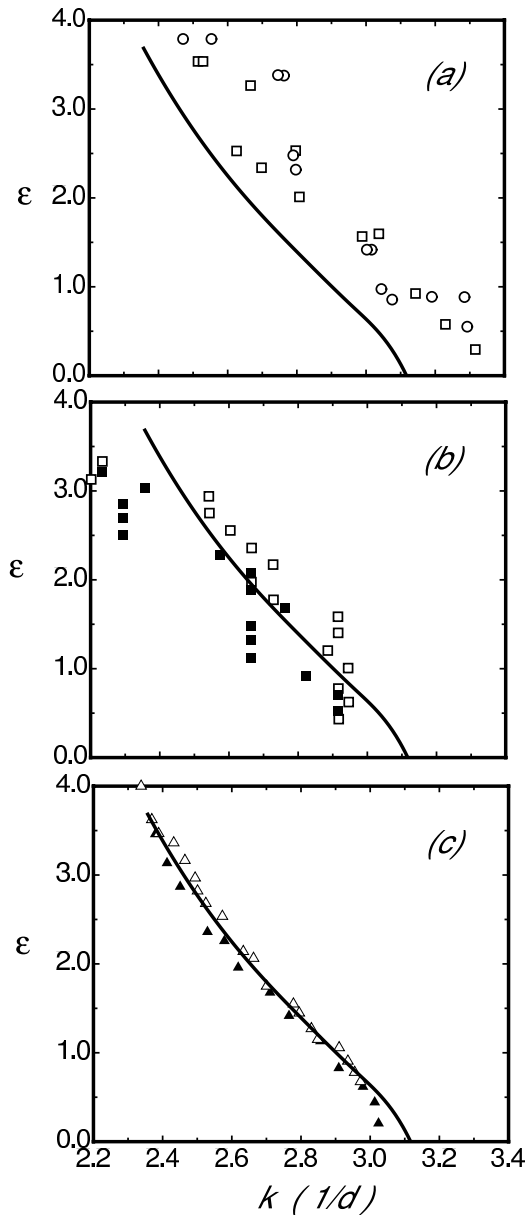}
\caption{Previous wave-number measurements for  patterns of concentric rolls. (a): from Ref.~\protect \cite{KP74} for $\sigma = $ 511 (circles) and 916 (squares), and for $\Gamma = 13.28$. For (b) and (c) open (solid) symbols were taken with increasing (decreasing) $\varepsilon$.  (b): from Ref.~\protect \cite{SAC85} for $\sigma = 6.1$ and $\Gamma=7.5$ .  (c): from Ref.~\protect \cite{CP84} for $\sigma=14$ and $\Gamma = 20$.  The solid lines in all figures are the predictions from Refs.~\protect \cite{BC86} and \protect \cite{NPS90} for the relevant $\sigma$ values.}
\label{fig:k_others}
\end{figure}

 For $\sigma$ values not too small the wave numbers predicted for curved rolls  do not depend very strongly on $\sigma$. In this $\sigma$ range the wave numbers of target patterns have been determined before, \cite{KP74,CP84,SAC85,CLPG86}  but to our knowledge they were never compared quantitatively with the predictions of BC. Koschmieder and Pallas \cite{KP74}  determined a weighted average of the wave lengths of all rolls except for the outermost one.  In Fig.~\ref{fig:k_others}a we show their results for $\Gamma = 13.28$, and for $\sigma = 511$ and 916. The solid line is the prediction of Ref.~\cite{BC86} for $\sigma = \infty$. Aside from a lateral shift, the overall shape of a curve passing through the data is very similar to the theoretical curve. Since the data do not extrapolate to $k_c = 3.117$ as $\varepsilon$ vanishes, the shift presumably is due to experimental uncertainties of the length scales involved in the determination of $\lambda$.

In Fig.~\ref{fig:k_others}b we show
results from Ref.~\cite{SAC85} which were for $\sigma=6.1$ and $\Gamma=7.5$. These data were obtained by measuring the average wave length of 2 rolls near $r \simeq 2\Gamma/3$. They seem to show considerable scatter, but actually this is due to the discreteness of the values of $\bar k$ which is noticeable in the figure and which becomes more apparent for this relatively small $\Gamma$ where the system contains only a small number of rolls. In this system hysteretic transitions were clearly observed. The  theoretical curve \cite{BC86} for this $\sigma$ value (solid line) is a good smoothed representation of the data. 
 
Finally, in Fig.~\ref{fig:k_others}c, some  of the results reported by Croquette and Pocheau ~\cite{CP84} are shown. Those measurements were for $\sigma=14$ and $\Gamma=20$, and were obtained by measuring the average roll width along a diameter in a region away from the center and sides where the roll width appeared constant.   The solid line is the numerical result \cite{BC86} for $\sigma = 14$. The agreement clearly is very satisfying. The data do not reveal much difference when $\varepsilon$ is increased (open symbols) or decreased (solid symbols). The authors report observing hysteretic transitions, and thus it is somewhat surprising that this hysteresis does not manifest itself in the wave-number measurements.

All of the previous investigations revealed that the roll adjacent to the side wall was exceptionally wide, but none of the prior investigations made any attempt to determine quantitatively the wave number selected at the wall or at the cell center.

\section{Experiment}
\label{sec:experiment}
\subsection{Apparatus}
\label{sec:apparatus}

We used a Rayleigh-B\'{e}nard convection apparatus described in detail elsewhere \cite{ACBS94,MA02}.  The top plate was  a single-crystal-sapphire disk, and the bottom plate was a polished aluminum disk with a metal-film heater mounted under it to provide uniform bottom-plate heating.  Temperature-controlled water was circulated over the top plate.     The bottom-plate temperature was controlled to create the desired temperature difference. The mean temperature was maintained at $22.0 ^{\circ}$C.  For both the top and bottom plate, the temperature varied about the set temperature by less then $10^{-3}\ ^{\circ}$C.

The cell wall was made of Lexan (thermal conductivity 0.23 W/m K) with an outer diameter of 9.6 cm and an inner diameter of $D = 8.89$ cm.    A 0.012 cm diameter manganin wire was embedded in the cell wall to provide the side-wall heating.  This wire had a resistance of about 13 $\Omega$,  and typically dissipated 0.26 W.

The height of the side wall was 0.229 cm, though the actual fluid height $d$ depended on the compression of an O-ring which sealed the cell.  The actual fluid height was determined by measuring $\Delta T_c$ and then inferring $d$ from the fluid properties and the measured $\Delta T_c = 0.873 \pm 0.005^{\circ}$C. We found $d=0.234$cm, giving an aspect ratio $\Gamma = 19.0$ .

The fluid used was methanol which, at $22.0^\circ$C,  had a Prandlt number $\sigma = 7.17$ and a thermal conductivity of $0.20$ W/m K.  The vertical diffusion time was $\tau_v=54$s.   

The convection rolls were imaged using the shadowgraph method \cite{BBMTHCA96,TC03}.  At large $\varepsilon$, strong spatial variation of the refractive index caused nonlinear optical effects in the images,\cite{TC03} but the general features of the pattern could still be discerned.   This effect limited our ability to do wave-number measurements for $\varepsilon \agt 4$. Nonetheless, the location of the umbilicus could still be measured.

In order to obtain concentric rolls, $\Delta T$ was set to zero and the side-wall heater-power was  set to 0.26 W.   After waiting two hours, we increased the temperature difference to $\Delta T =0.78^\circ$C ($\varepsilon \simeq -0.1$) and allowed the system to equilibrate for one hour. After this the temperature difference was abruptly increased to $\Delta T = 1.3^\circ$C ($\varepsilon \simeq 0.5$) and the system was allowed to equilibrate for two hours.  The jump of $\Delta T$ was necessary because defects typically formed in the interior when the temperature difference was raised gradually.   Once concentric rolls were obtained, the temperature difference was adjusted in small steps (typically between $0.05^\circ$C and $0.01^\circ$C) and the system equilibrated for one hour before taking an image at each temperature step.  The power supplied by the side-wall heater was kept constant throughout the run.  

\subsection{Image Analysis}
\label{sec:image_analysis}

\begin{figure}
 \includegraphics[width=7cm]{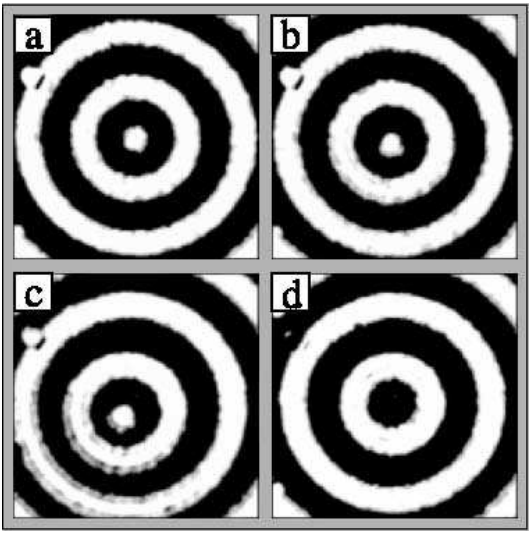}
\caption{The central $11.4d\times 11.4d$ section of the pattern for $\varepsilon$ values near the transition at $\varepsilon_{4,b}$. They are from a run in which $\Delta T$ was increased with $0.05^{\circ}$C steps $(\Delta\varepsilon = 0.057)$. The system was equilibrated for one hour after each temperature step before taking a picture. The images were divided by a background image, and then processed numerically so as to greatly enhance their contrast.  They are for (a) $\varepsilon = 3.139$, (b) $\varepsilon = 3.244$, (c) $\varepsilon = 3.301$, and (d) $\varepsilon = 3.359$. }
\label{fig:transition_up}
\end{figure}
 
\begin{figure}
 \includegraphics[width=7cm]{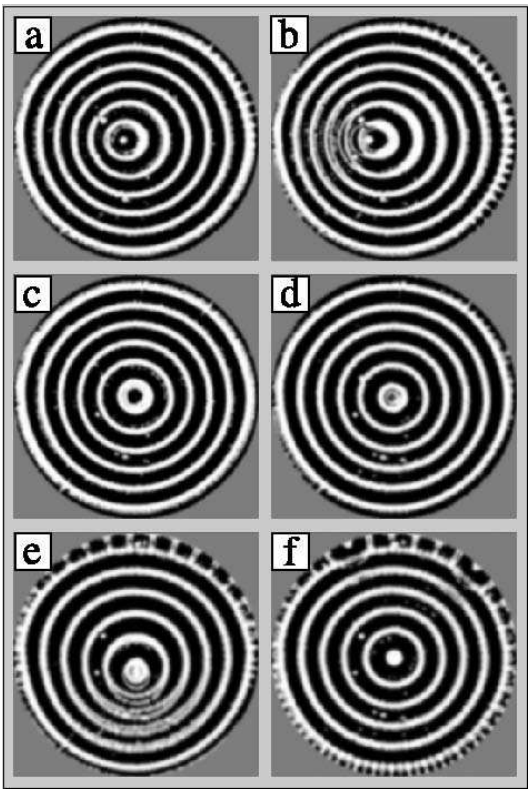}
\caption{Images near the transitions at $\varepsilon_{6,b}$ and $\varepsilon_{7,b}$.  The optical effects that distort the roll patterns can be seen in images (b) and (e) in the rolls that are being compressed by the off-center umbilicus.  These images are from the same run as those in Fig.~\ref{fig:transition_up}. The values of $\varepsilon $ are (a) 5.649, (b)  5.707, (c)  5.821, (d)  6.910, (e) 7.368, and (f)  7.482. }
\label{fig:transition_higher}
\end{figure}

Shadowgraph images of some of the patterns are shown in Figs.~\ref{fig:transition_up} and \ref{fig:transition_higher}.  All images were divided by a background image taken at $\Delta T = 0$.
They were filtered by setting to black (white) all pixels that fell below (above) a certain threshold. A gaussian blur was then applied to the chopped image.  
  
We developed an algorithm to determine the local wave-director field $\vec k$.  This method will be described in detail elsewhere.  First the orientation field $\theta (x,y) $ at each point of the image was calculated using a method similar to one introduced by Cross et. al., \cite{CMT94} but higher angular resolution was employed. The wave-number field $k(x,y)$ was then calculated at each point by moving orthogonal to the the roll orientation at that point and determining the number of pixels which falls within one wave length, assuming a locally periodic structure.

Determinations of the wave number $2 \pi d / \lambda$ from the images require a knowledge of $d$ and of the horizontal distance $\delta x$ between adjacent pixels. The uncertainty of $d$ yielded an uncertainty of only a small fraction of a percent.  We determined $\delta x$ by counting the number of pixels which spanned the sample diameter.
This yielded $\delta x = 0.095 \pm .002$ and led to a combined uncertainty of about 2 \% for $k$. An extrapolation of the measured $k$-values with decreasing $\varepsilon$ to $\varepsilon = 0$ gave $k_c = 3.166$, which is within the experimental uncertainty of the value 3.117 for the infinitely extended set of straight rolls.  

We measured the location of the umbilicus by using a $24d \times 24d$ square section surrounding the center.  The location of the umbilicus was estimated by computing the maximum of the cross-correlation of the measured orientation field $\theta(x,y)$ and the orientation field for perfectly centered rolls $\theta_c (x,y)$.  This estimate was used only to identify the rough location of the umbilicus. Once the umbilicus was identified, we used an edge-detection algorithm on the umbilicus to determine its position with a resolution of about $ 70 \mu$m or $0.03d$.
 
\section{Results}
\label{sec:results}

\subsection{Patterns: Increasing $\varepsilon$}
\label{sec:patterns_up}

We covered the range $0< \varepsilon <15.6 $ and with increasing $\varepsilon$ observed eight transitions at $\varepsilon_{i,b}, i = 1, ..., 8$ under quasi-static conditions. A closeup view of images near the third transition is shown in Fig.~\ref{fig:transition_up}.    
At each transition  a roll was lost at the umbilicus. For all but the first transition the umbilicus first moved away from the center as $\varepsilon$ grew.  The angle of the umbilicus displacement did not have a preferred value, as can be seen from the examples in Figs.~\ref{fig:transition_up} and \ref{fig:transition_higher}. This  indicates that the cell was sufficiently uniform. At the transition the umbilicus collapsed, reducing the number of rolls in the system by one. After a roll was lost, the new umbilicus immediately moved back toward the center of the sample but never quite reached it.  
This process was analogous to that observed in Refs.~\cite{CP84} and \cite{SAC85}.  
The umbilicus displacement before each transition produced a noticeable azimuthal variation of the wave number, with higher wave numbers in the direction of the displacement, and lower wave numbers opposite the displacement direction. 

Close to but before  the sixth transition, for $\varepsilon \simeq 5.4$, a patch of cross rolls \cite{CB74,BC79} formed along the outermost roll on the side opposite the direction of umbilicus displacement.  At that point the local wave number $k$ of the concentric rolls was the smallest of the entire wave-number field, and was about 1.95 ($k/k_c - 1 = -0.39$). The corresponding point in the $\varepsilon-k$ plane is shown as a solid circle in Fig.~\ref{fig:bussball}.  The wave number $b$ of the cross rolls that formed was about 4.3 .  The value of $k$ is somewhat larger than the prediction \cite{BC79} $k_{CR} = 1.67$ ($k_{CR}/k_c - 1 = -0.46$) for laterally infinite straight rolls at this value of $\varepsilon$. This suggests that the roll curvature or the side wall reduces the stablity of the rolls against the cross-roll perturbation. The wave number $b$ of the cross rolls that formed is larger than  the predicted value \cite{BC79} $b_{CR} \simeq 3.5$. The cross rolls disappeared when $\varepsilon$ was increased further, the pattern lost the middle roll, and the umbilicus returned toward the center. At that point all parts of the pattern had returned to wave numbers safely in the stable Busse balloon. Before the next (seventh) transition, near $\varepsilon \simeq 6.9$, cross rolls appeared again. In this case we found $k$ =1.86 ($k/k_c - 1 = -0.41$) and $b = 4.3$. This point is shown as well in Fig.~\ref{fig:bussball}. Again the value of $b$ is somewhat larger than the prediction $b \simeq 3.9$ for infinitely extended uniform straight rolls. After this transition the cross rolls fanned out to cover the entire outer roll.  Images near the two transitions are shown in Fig.~\ref{fig:transition_higher}.   

\begin{figure}
 \includegraphics[width=8.5cm]{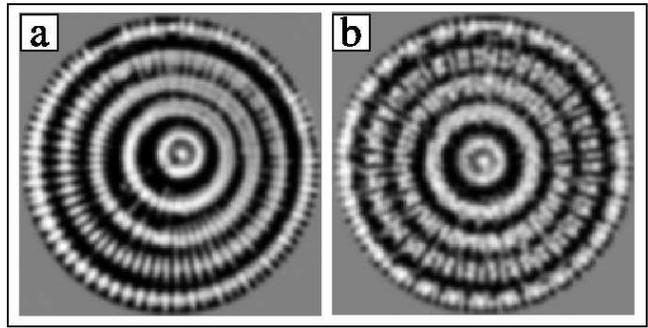}
\caption{Images for (a) $\varepsilon = 12.75 $ and (b) $\varepsilon = 15.56$ from the same run as the one used for Figs. \ref{fig:transition_up} and \ref{fig:transition_higher}).  The pattern in image (a) was stationary, while in image (b) the cross rolls oscillated along the axis of the main rolls. At these large $\varepsilon$ values the images are strongly influenced by nonlinear effects in the shadowgraph method.}
\label{fig:large_eps}
\end{figure}

The cross rolls remained along the outer roll, and after the eighth transition, spread to the inner rolls as $\varepsilon$ was further increased.  Despite the cross rolls, there was still a discernible pattern of nearly concentric rolls, with the cross rolls superimposed on these.     Around $\varepsilon \simeq 11.5$  the focus began to move off center  but did not complete a ninth transition. This is illustrated in Fig.~\ref{fig:large_eps}a for $\varepsilon = 12.75$. Cross rolls spread to the inner rolls, and as $\varepsilon$ was further increased, the umbilicus returned to the center without losing a roll, as seen in Fig.~\ref{fig:large_eps}b for $\varepsilon = 15.56$.

The pattern became time dependent near $\varepsilon = 14.2$,  where small patches of traveling waves developed. These were superimposed upon and orthogonal to the concentric roll, and  traveled along the axes of the concentric rolls. The travelling waves occurred first on the third or fourth roll from the wall, and with increasing $\varepsilon$ spread throughout the cell. We asociate this phenomenon with the oscillatory instability predicted by Clever and Busse, \cite{CB74,BC79} although for our Prandtl number  the cross-roll instability precedes the oscillatory instability.

The cross rolls observed here had been previously seen by Croquette et. al. \cite{CP84}, though there are differences in the way they appeared.  Croquette et. al. observed that cross rolls moved in from the side wall only after $\varepsilon$ became larger than about 10.  They did not observe a relationship between the umbilicus transitions and the cross-roll formation. Despite nearly identical Prandlt numbers and aspect ratios ($\sigma=7$, $\Gamma=20$), Croquette et. al. could produce off-center patterns  near transitions comparable to those shown in Fig.~\ref{fig:transition_higher} with no cross rolls \cite{CGP86}.   In our system cross rolls permanently covered the outer roll for $\varepsilon=7.48$, well before Croquette et. al. first observed any. 
This difference can perhaps be explained by a difference  in the strength of  side-wall forcing. A future study of the influence of various levels of side-wall heating should shed some light on this issue.

\begin{figure}
 \includegraphics[width=7cm]{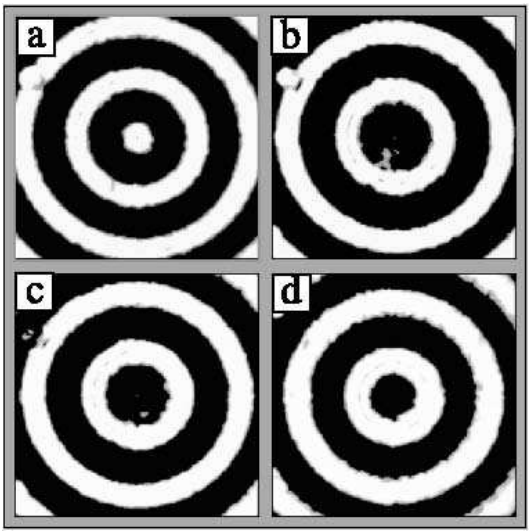}
\caption{The central $11.4d\times 11.4d$ sections of images obtained with decreasing $\varepsilon$ near the transition at $\varepsilon_{4,a}$. This is from a run where $\varepsilon$ was first increased to 3.49 and then decreased in steps $\Delta\varepsilon = -0.01$.The values of $\varepsilon$ are (a) 3.49, (b) 2.94, (c) 2.67, and (d) 2.65.  Note how the size of the middle roll changes from image (a) to (c), and how in (c) the umbilicus remains close to the center even for an $\varepsilon$ value only just above $\varepsilon_{4,a}$.}
\label{fig:transition_down}
\end{figure}

\subsection{Patterns: Decreasing $\varepsilon$}
\label{sec:patterns_down}

The nature of the patterns for decreasing $\varepsilon$ depended on the initial state.
When starting above the oscillatory instability (say $\varepsilon \agt 14$), decreasing $\varepsilon$ broke up the concentric rolls.   Wall foci evolved, displaced the umbilicus, and emitted new rolls as $\varepsilon$ was further decreased.  However, the side-wall forcing was strong enough to re-orient the outermost rolls parallel to the sidewall for  $\varepsilon \alt 0.21$.   By $\varepsilon \simeq 0.09$ the outer roll had returned to being completely parallel to the sidewall.  However, defects remained in the interior of the pattern, and the system did not return to concentric rolls as $\varepsilon$ was decreased below zero.

Starting from $\varepsilon \alt 6$ with concentric time-independent rolls, new rolls were generated at the umbilicus for distinct values $\varepsilon_{i,a} < \varepsilon_{i,b}$ as $\varepsilon$ was decreased quasi-statically. During these transitions the umbilicus stayed close to the center, as illustrated in Fig.~\ref{fig:transition_down}.   
We did not determine the maximum $\varepsilon$ at which we could start decreasing $\varepsilon$ without loosing the concentric pattern.  

\begin{figure}
 \includegraphics[width=7cm]{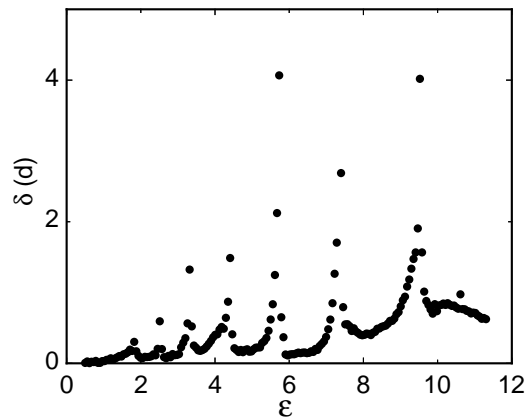}
\caption{The umbilicus displacement $\delta$ away from the sample center as a function of $\varepsilon$ for increasing $\varepsilon$. This data is from the same run as the images in Figs.~\ref{fig:transition_up} and \ref{fig:transition_higher}, where we increased $\varepsilon$ in steps of $\Delta\varepsilon = 0.057$ and waited 1 hour after each temperature step for the system to equilibrate before taking a picture. }
\label{fig:dist_umbil}
\end{figure}

\subsection{Umbilicus Displacement}
\label{sec:umbil_dis}

\begin{figure}
 \includegraphics[width=7cm]{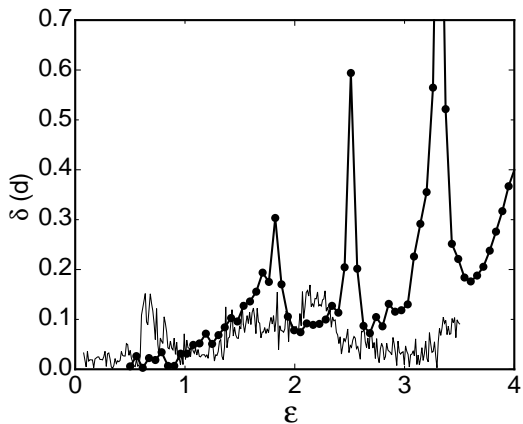}
\caption{An expanded view at relatively small $\varepsilon$ of the results shown in Fig.~\ref{fig:dist_umbil}.  The thin solid line shows the umbilicus displacement for decreasing $\varepsilon$.   }
\label{fig:dist_umbil_detail}
\end{figure}

With increasing $\varepsilon$ we measured the displacement $\delta$ of the umbilicus from the center of the cell as a function of $\varepsilon$ for $0.5<\varepsilon<12$.  Results over the entire $\varepsilon$-range are shown in Fig.~\ref{fig:dist_umbil}. A more detailed plot of the results for $\varepsilon < 4$ is shown  in Fig.~\ref{fig:dist_umbil_detail}. Seven transitions, at $\varepsilon_{i,b}, i = 2, ..., 8$, are apparent from these data. At the first transition, at $\varepsilon_{1,b}$, we did not resolve any discontinuity in $\delta$. Prior to each observable transition there was a large displacement, followed by a relaxation  back toward a centro-symmetric pattern when the umbilicus collapsed. 

\begin{figure}
 \includegraphics[width=7cm]{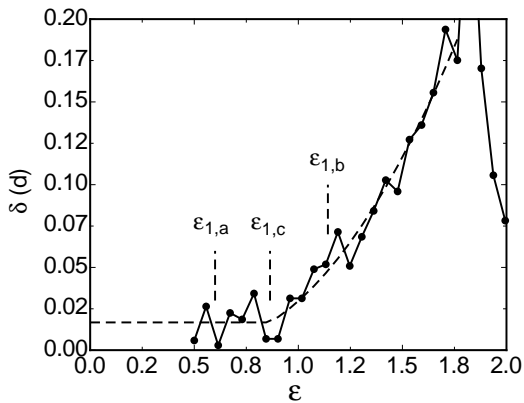}
\caption{A very expanded view at small $\varepsilon$ of the results shown in Figs.~\ref{fig:dist_umbil} and \ref{fig:dist_umbil_detail}. This graph reveals the initial focus instability of the centro-symmetric pattern at $\varepsilon_{1,c}$. The points $\varepsilon_{1,a}$ and $\varepsilon_{1,b}$ are the limits of the hysteresis loop associated with the transition which are revealed by  the wave-number measurements to be discussed below. The dashed curve is a fit of Eq.~\ref{eq:delta(eps)} to the data.}
\label{fig:dist_umbil_detail2}
\end{figure}

Figure~\ref{fig:dist_umbil_detail2} shows the data for $\delta$ over the range $\varepsilon < 2$. Here we see that the pattern is, within possible systematic errors of $\delta$, accurately centered for $\varepsilon < \varepsilon_{1,c} \simeq 0.8$. Above $\varepsilon_{1,c}$ the increase of $\delta$ is consistent with an initially linear dependence, and thus we fitted the results to 
\begin{equation}
\delta(\varepsilon) = \delta_0 + \delta_1 (\varepsilon - \varepsilon_{1,c}) + \delta_2 (\varepsilon - \varepsilon_{1,c})^2
\label{eq:delta(eps)}
\end{equation}
 over the range $\varepsilon_{1,c} \leq \varepsilon \leq 1.7$ and to $\delta(\varepsilon) = \delta_0$ for $\varepsilon < \varepsilon_{1,c}$. Here $\delta_0$ corresponds to the small offset, well within our possible systematic errors, which is found even at small epsilon. The parameters $\delta_0, \delta_1, \delta_2$, and $\varepsilon_{1,c}$ were least-squares adjusted. The fit gave $\varepsilon_{1,c} = 0.86 \pm 0.13$, $\delta_1 = 0.077 \pm 0.075$, and $\delta_2 = 0.12 \pm 0.05$. It is shown as the dashed curve in Fig.~\ref{fig:dist_umbil_detail2}. The statistical error of $\delta_1$ indicates that a fit to a quadratic equation (i.e. Eq.~\ref{eq:delta(eps)} with $\delta _1 = 0$) should be equally good. It yielded $\varepsilon_{1,c} = 0.66 \pm 0.08$ and $\delta_2 = 0.14 \pm 0.02$. We identify $\varepsilon_{1,c}$ as the focus instability, i.e. as the first instability of the centro-symmetric pattern with our prevailing wave number and our Prandtl number $\sigma = 7.17$. 

When $\varepsilon$ was decreased, transitions involving the addition of a roll at the umbilicus occurred at $\varepsilon_{i,a}$, but were not associated with large displacements of the umbilicus. Experimental results for $\delta$ with decreasing $\varepsilon$ are shown by the thin solid line in Fig.~\ref{fig:dist_umbil_detail}. This difference between increasing and decreasing $\varepsilon$ was observed also by Steinberg et. al. \cite{SAC85}. 

\subsection{Wave-Number Measurements}
\label{sec:wavenumbers}

\begin{figure}
 \includegraphics[width=6cm]{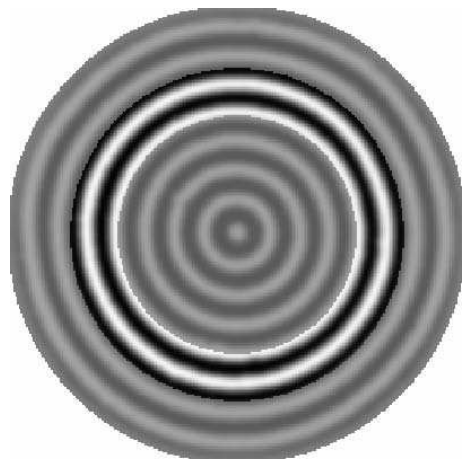}
\caption{A chopped and blurred image obtained with increasing $\varepsilon$ at $\varepsilon = 2.61$.  The highlighted section covers the range $10 \leq r \leq 14$ and shows the annulus over which we averaged the wave numbers to obtain the values $\langle k \rangle$ shown in Figs.~\ref{fig:eps_k_down}, \ref{fig:eps_k}, and \ref{fig:eps_k_detail} below.}
\label{fig:image}
\end{figure}

\begin{figure}
 \includegraphics[width=7cm]{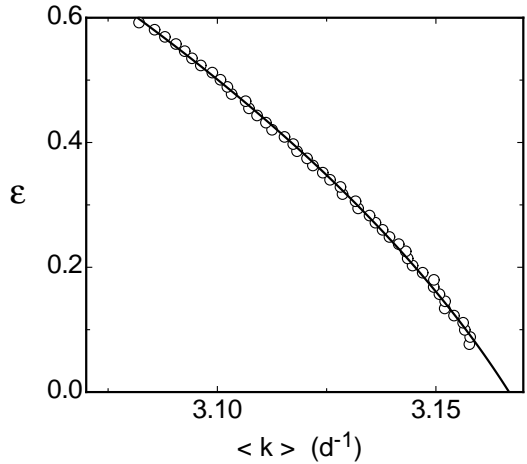}
\caption{Results for $\langle k \rangle$ obtained with decreasing $\varepsilon$. The solid line is a fit of Eq.~\ref{eq:k_B} to the data.}
\label{fig:eps_k_down}
\end{figure}

We measured the averaged wave numbers $ \langle k \rangle$ for $0<\varepsilon<4$ for both increasing and decreasing $\varepsilon$. For comparison with theoretical predictions  one would like to make this measurement well away from the umbilicus;  in practice there is a limit set by the aspect ratio of the sample and by a competing selection mechanism associated with the side wall. As we shall discuss below, the local wave numbers have a maximum at a radial position in the range $10 < r < 14$ ($0.53  \Gamma < r < 0.74 \Gamma$). Thus we selected an annular region extending over this radial range  as illustrated by the highlighted region in Fig.~\ref{fig:image} to compute $\langle k \rangle$.
In Fig.~\ref{fig:eps_k_down} the results obtained with decreasing and relatively small $\varepsilon$ are shown. The solid line is a fit of the equation
\begin{equation}
\langle k \rangle /k_c - 1 = S_B \varepsilon + S_2 \varepsilon^2
\label{eq:k_B}
\end{equation}
to the data over the range $\varepsilon < 0.6$.  The fit yielded $k_c = 3.167$, $S_B = -0.0285 \pm 0.0010$, and $S_2 = -0.027$. The result for $k_c$, within the a priori estimate of the experimental uncertainties, agrees with the theoretical value $k_c = 3.117$ for infinitely extended straight rolls. The result for $S_B$ is shown in Fig.~\ref{fig:S_B} as an open circle. It falls slightly above the predicted value $-0.0317$.

\begin{figure}
 \includegraphics[width=8.5cm]{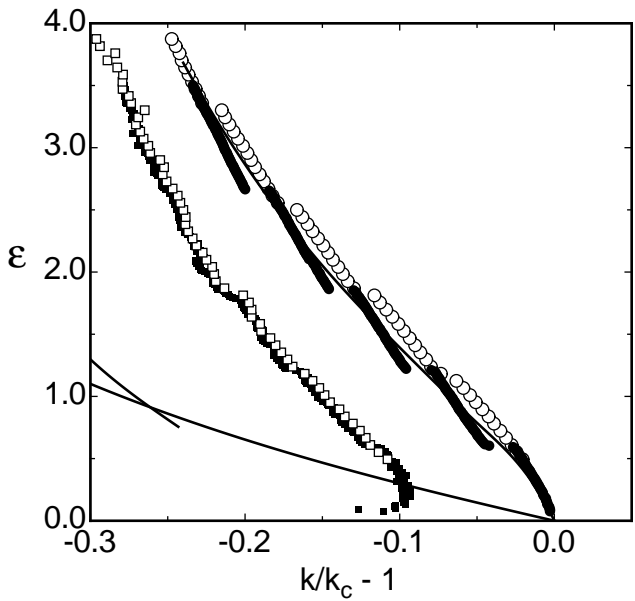}
\caption{Circles: the average selected reduced wave numbers $\langle k \rangle / k_c - 1$ vs. $\varepsilon$. Squares: the reduced wave numbers $k_\theta(\Gamma) / k_c - 1$ selected at the side wall. Open symbols: increasing $\varepsilon$. Closed symbols: decreasing $\varepsilon$.  Solid line through the data for $\langle k \rangle/k_c - 1$: the prediction of Refs. \cite{BC86} and \cite{NPS90}. Lower solid line: the zigzag instability for $\sigma=7$. }
\label{fig:eps_k}
\end{figure}

Results for $\langle k \rangle / k_c - 1$ over the entire $\varepsilon$ range are shown in Fig.~\ref{fig:eps_k} for both increasing (open circles) and decreasing (solid circles) $\varepsilon$.  Within experimental uncertainties they are consistent with previous measurements for similar $\sigma$. \cite{SAC85, CP84}.  The jumps in the wave number correspond to the loss or creation of a roll at a transition.   The hysteresis in the wave-number selection, previously observed by others, \cite{SAC85, CP84} is clearly visible in our results. It indicates partial pinning of the phase of the pattern at the umbilicus. We believe that this phase pinning is responsible also for the difference between experiment and theory for the value of $S_B$. On average the data are in excellent agreement with the calculations by Buell and Catton \cite{BC86} which are given by the solid line through the data.

\begin{figure}
 \includegraphics[width=8.5cm]{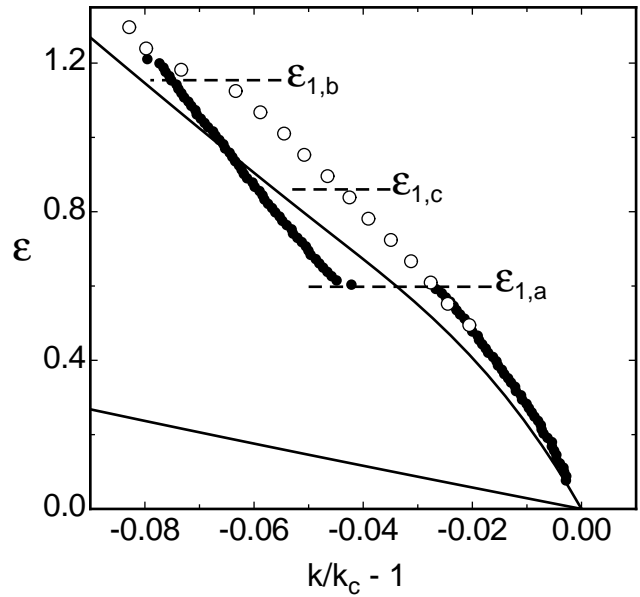}
\caption{A detailed view of selected wave numbers $\langle k \rangle$ near the focus instability at $\varepsilon_{1,c}$ and the first hysteretic transition at $\varepsilon_{1,b}$ and $\varepsilon_{1,a}$. The symbols are as in Fig.~\ref{fig:eps_k} }.
\label{fig:eps_k_detail}
\end{figure}   

In order to illustrate the relationship between the first hysteretic transition at $\varepsilon_{1,a}$ and $\varepsilon_{1,b}$ and the focus instability at $\varepsilon_{1,c}$,  
we give an expanded view of the selected wave numbers at small $\varepsilon$ in Fig.~\ref{fig:eps_k_detail}. One sees that the discontinuities at $\varepsilon_{1,a}$ and $\varepsilon_{1,b}$ are quite sharp, and that $\varepsilon_{1,c}$ is located near the middle of this first hysteresis loop. Values of $\varepsilon_{i,a}$ and $\varepsilon_{i,b}$ for all eight transitions are given in Table~\ref{table:eps}.

\begin{table}
\caption{Location of the hysteretic transitions with decreasing ($\varepsilon_{i,a}$) and increasing ($\varepsilon_{i,b}$) $\varepsilon$. For $\varepsilon_{i,a}$, $i=1,..,4$ the uncertainty $\delta \varepsilon = 0.006$.  For $\varepsilon_{5,a}$ and $\varepsilon_{6,a}$, $\delta \varepsilon =  0.11$.  For $\varepsilon_{i,b}$, $\delta \varepsilon = 0.029$.  }
\vskip 0.1in
\begin{tabular}{ccc}
$ i $ & $\varepsilon_{i,a}$ &$\varepsilon_{i,b}$\\
\colrule
1	& 0.598	& 1.154\\
2	& 1.217	& 1.841 \\
3	& 1.858	& 2.528\\
4	& 2.660	& 3.330\\
5  	& 3.6	& 4.437\\
6  	& 4.8	& 5.759 \\
7	&  -		& 7.368 \\
8	&  -		& 9.552\\
\end{tabular}
\label{table:eps}
\end{table}

\begin{figure}
 \includegraphics[width=8.5cm]{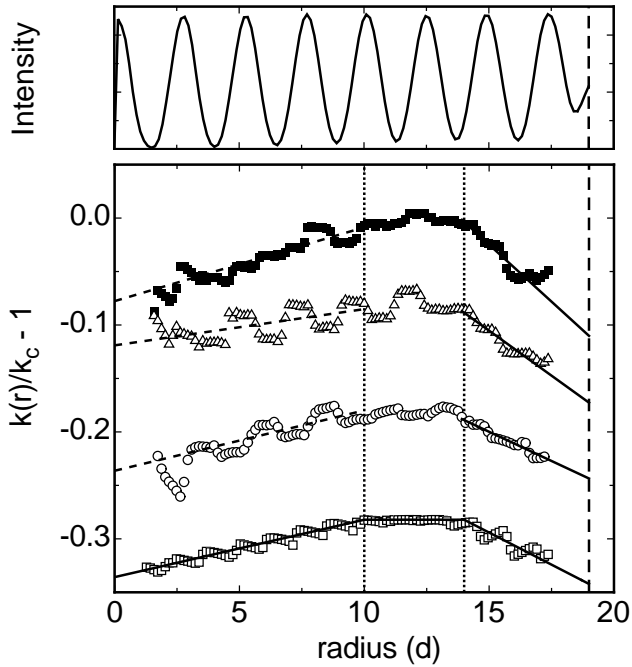}
\caption{Top: an azimuthal  average of the shadowgraph intensity (arbitrary scale) as a function of radial position for $\varepsilon = 2.61$ . Bottom: azimuthal averages of the local wave numbers.  The vertical dashed line shows the location of the side wall, and the two vertical dotted lines show the inner and outer edge of the annulus highlighted in Fig.\ref{fig:image} and used to compute the average wave numbers $\langle k \rangle$ shown in Figs.~\ref{fig:eps_k_down}, \ref{fig:eps_k}, and \ref{fig:eps_k_detail}. From top to bottom, the first three data sets are for $\varepsilon = 0.077$, 1.297, and 2.614. The lowest set is for a synthetic concentric pattern (down-shifted in the figure by 0.1) with a wave-number distribution (shown by the solid line) given by straight-line representations of the experiment for $\varepsilon = 2.614$. The solid straight lines through the experimental data at large $r$ are fits to the data with $r \geq 14$ and were used to find $k_\theta(\Gamma)$. The dashed straight lines through the experimental data at small $r$ are fits for $5 \leq r \leq10$ and were used to determine $k(0)$.}
\label{fig:k_rad}
\end{figure}

\begin{figure}
 \includegraphics[width=8.5cm]{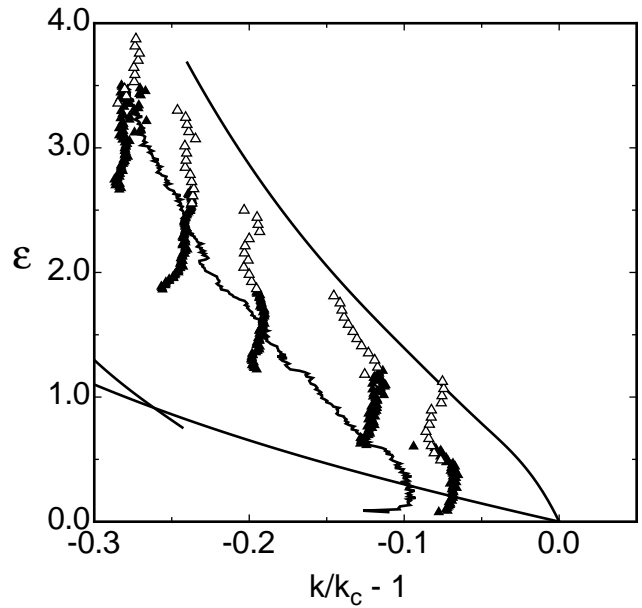}
\caption{Triangles: The reduced wave numbers $k(0)/k_c - 1$ selected at the origin.  Open symbols: increasing $\varepsilon$. Closed symbols: decreasing $\varepsilon$.  Rough solid line: $k_\theta(\Gamma)$ for decreasing $\varepsilon$. Smooth solid lines: predictions for $\langle k \rangle$ \protect \cite{BC86} and for the zigzag and cross-roll instabilities \protect \cite{BC79}.}
\label{fig:k_0}
\end{figure}

The top part of Fig.~\ref{fig:k_rad} gives an example for the radial variation of the shadowgraph intensity. In the bottom part 
we display the radial variation of the azimuthal average $k_\theta$ of $k$ for three $\varepsilon$ values. Also shown for comparison (bottom curve, down-shifted by 0.1) is the result of our analysis for a synthetic pattern of concentric rolls with a radial wave-number distribution shown by the solid line which represents a smooth curve through the results for $\varepsilon = 2.614$. 

Our analysis yields results for $k_\theta(r)$ which oscillate as a function of $r$. This is particularly noticeable for $r$ below the maximum of $k_\theta(r)$. These oscillations are not present in the analysis of a synthetic concentric pattern with a uniform wave number, but as shown by the lowest set  in Fig.~\ref{fig:k_rad}, they do appear in the analysis of a concentric pattern with a radial wave-number distribution equal to the smooth curve through the data for $\varepsilon = 2.614$. Thus we regard them to be an artifact of the numerical procedures. We note, however, that the amplitude of the oscillations is only about 0.5\% of $k$, and that a smooth curve through them is a good representation of the actual wave numbers.  

From Fig.~\ref{fig:k_rad} one sees that the experimental data for $k_\theta$, as mentioned above, have a maximum in the range $10 < r < 14$, and this range was used to compute $\langle k \rangle$. For $r > 14$ the data were fitted by straight lines as shown in the figure, and these fits were used to extrapolate the selected $k_\theta(r)$ to $k_\theta(r = \Gamma)$. The results for $k_\theta(\Gamma)$ are shown in Fig.~\ref{fig:eps_k} above. They show, as noted by others, \cite{KP74,CP84,SAC85} that the outermost roll has an anomalously large wave length (small wave number). We note that $k_\theta(\Gamma)$ is not influenced significantly by the transitions at the umbilicus. Interestingly, the wall-selected wave number does not extrapolate to $k_c$ as $\varepsilon$ vanishes. Instead, it crosses the zigzag instability-line of the laterally infinite system; but in the experiment for this finite system no instability was observed. We are not aware of any quantitative previous measurements or predictions for $k_\theta(\Gamma)$.

The experimental results at small $r$ suggest that yet another unique wave number is selected by the umbilicus itself. Fits of straight lines to the data, this time with $5 < r < 10$, could be used to estimate the selected values $k(0)$ at $r = 0$. The results are given in Fig.~\ref{fig:k_0}. For comparison we show  in that figure also the prediction for the concentric-roll selection \cite{BC86} (smooth solid line) and the experimental results for $k_\theta(\Gamma)$ (jagged line). One sees that $k(0)$ is influenced strongly by the hysteretic transitions at the umbilicus. At large $\varepsilon$ $k(0)$ is on average close to the result for $k_\theta(\Gamma)$, but as $\varepsilon$ decreases, $k(0)$ is somewhat larger and tends toward an intermediate value somewhere between $k_\theta(\Gamma)$ and $\langle k\rangle$ as $\varepsilon \rightarrow 0$. Also for $k(0)$ there appear to be no previous measurements or predictions.

\section{Conclusion}
\label{sec:conclusion}
 
 We studied Rayleigh-B\'{e}nard convection of a fluid with a Prandtl number $\sigma = 7.17$  in a cylindrical container of aspect ratio $\Gamma = 19$ in the presence of static side-wall forcing.  Patterns of concentric rolls were obtained, and we studied the quasi-static evolution of these patterns for both increasing and decreasing $\varepsilon$.  
 
 For increasing $\varepsilon$, over a range $0.4<\varepsilon<15.6$, the pattern underwent eight transitions where the middle roll moved off center and then disappeared as $\varepsilon$ was increased quasi-statically.  We measured the displacement of the umbilicus for $0.5 < \varepsilon<12$.   Above $\varepsilon\simeq14$ the pattern became time dependent with oscillating rolls traveling along the axis of the concentric rolls.    
   
 For decreasing $\varepsilon$, the concentric-roll pattern was lost when the initial $\varepsilon$ was too large.  When  $\varepsilon$ was decreased from a moderate initial value, the concentric-roll pattern remained and new rolls nucleated at the umbilicus. Any umbilicus displacement preceding the nucleation of new rolls was modest, in sharp contrast with the behavior of the pattern for increasing $\varepsilon$.
 
Using new image-analysis techniques, we determined  the wave-number field for $0 < \varepsilon < 4$.  From this we computed the azimuthal average $k_\theta$. There was a very noticeable radial gradient of $k_\theta$ both near the side wall at $r = \Gamma$
 and near the center at $r = 0$.  Averaging the wave-number field over the annulus $10 < r < 14$ where $k_\theta$ was relatively constant, we obtained an average wave number $\langle k \rangle$. We found that $\langle k \rangle$ was discontinuous and hysteretic at the transitions where new rolls were formed or disappeared at the umbilicus.
 Nonetheless, a smooth curve through the data agreed well with predictions \cite{BC86,NPS90} for the wave-number selection in the far field of concentric rolls. These average wave numbers were also consistent with previous work by others. 
 \cite{SAC85,CP84} Near the wall and near the center the measured $k_\theta$ were extrapolated to $r = \Gamma$ and to $r = 0$ to obtain the wave numbers $k_\theta(\Gamma)$ and $k(0)$ selected by the wall and the umbilicus. We are not aware of previous measurements or predictions for these quantities.
 
 \section{Acknowledgment}
\label{sec:acknowledgment}

We are grateful to Worawat Meevasana for his guidance during the early stages of this experiment, and to W. Pesch for stimulating discussions. This work was supported by US National Science Foundation Grant DMR02-43336.


\begin{thebibliography}{}

\bibitem{Ch61} For a review of stability analysis of RBC, see for instance S. Chandrasekhar, {\it Hydrodynamic and Hydromagnetic Stability}, (Oxford University Press, London, 1961).

\bibitem{SLB65}A. Schl\"uter, D. Lortz, and F. Busse, J. Fluid Mech. {\bf 23}, 129(1965).

\bibitem{BPA00}See, for instance, E. Bodenschatz, W. Pesch, and G. Ahlers, Annu. Rev. Fluid Mech. {\bf 32}, 709 (2000).

\bibitem{FNrolls}Perhaps the most convincing evidence for parallel straight rolls is found in the experiments of Ref.\cite{BMCA99} where the effect of side walls on the pattern was minimized by introducing a radial ramp of the cell spacing near the cell wall which gradually reduced the convection amplitude to zero as the wall was approached.

\bibitem{BMCA99}K.M.S. Bajaj, N. Mukolobwiez, N. Currier, and G. Ahlers, Phys. Rev. Lett. {\bf 83}, 5282 (1999).

\bibitem{KP74} L. Koschmeider and S. Pallas, J. Heat Mass Trans. {\bf 17}, 991 (1974).

\bibitem{CP84} V. Croquette, A. Pocheau. {\it Cellular Structures and Instabilities}, edited by J. E. Wesfried and S. Zaleski (Springer-Verlag, Berlin, 1984), p. 104.

\bibitem{SAC85} V. Steinberg, G. Ahlers, and D. Cannell, Phys. Scr. {\bf 32}, 534 (1985). 

\bibitem{CLPG86}V. Croquette, P. Le Gal, A. Pocheau, and R. Guglielmetti, Europhys. Lett. {\bf 1}, 393 (1986).

\bibitem{CGP86} V. Croquette, P. Le Gal, and A. Pocheau, Europhys. Lett. {\bf 1}, 393 (1986).

\bibitem{HEA93} Y. Hu, R. Ecke, G. Ahlers, Phys. Rev. E {\bf 48}, 4399 (1993). 

\bibitem{NPS90}A.C. Newell, T. Passot, and M. Souli, Phys. Rev. Lett. {\bf 64}, 2378 (1990); J. Fluid Mech. {\bf 220}, 187 (1990).

\bibitem{CB74}R.M. Clever and F.H. Busse, J. Fluid Mech. {\bf 65}, 625 (1974).

\bibitem{BC79}F.H. Busse and R.M. Clever, J. Fluid Mech. {\bf 91}, 319 (1979).

\bibitem{HG87}M.S. Heutmaker and J.P. Gollub, Phys. Rev. A {\bf 35}, 242 (1987).

\bibitem{CMT94} M. Cross, D. Meiron, and Y. Tu, Chaos {\bf 4},  607 (1994). 

\bibitem{EMB98}D.A. Egolf, I.V. Melnikov, and E. Bodenschatz, Phys. Rev. Lett. {\bf 80}, 3228 (1998).

\bibitem{TB88}L.S. Tuckerman and D. Barkley, Phys. Rev. Lett. {\bf 61}, 408 (1988).

\bibitem{BT89}D. Barkley and L.S. Tuckerman, Physica D {\bf 37}, 288 (1989).

\bibitem{PM81}Y. Pomeau and P. Manneville, J. Phys. (Paris) {\bf 42}, 1067 (1981).

\bibitem{Cr83} M.C. Cross, Phys. Rev. A {\bf 27}, 490 (1983).

\bibitem{MP83}P. Manneville and J.M. Piquemal, Phys. Rev. A {\bf 28}, 1774 (1983).

\bibitem{BC86} J. Buell and I. Catton, Phys. Fluids {\bf 29}, 1 (1986).

\bibitem{CDHS} M.C. Cross, P.G. Daniels, P.C. Hohenberg, and E.D. Siggia, Phys. Rev. Lett. {\bf 45}, 898 (1980); J. Fluid Mech. {\bf 127}, 155 (1983).

\bibitem{FNTW2}Traveling waves induced by two competing selection mechanisms have been studied \cite{RBWB87} in a one-dimensional system terminated at the two ends by two different kinds of ramps of the Rayleigh number.

\bibitem{RBWB87} I. Rehberg, E. Bodenschatz, B. Winkler, and F. H. Busse, Phys. Rev. Lett. {\bf 59}, 282 (1987). 

\bibitem{FNTW}Inward-traveling concentric rolls have been reported in Ref.~\cite{TBA02}, but for a probably unrelated case of convection in the presence of rotation about a vertical axis with periodic modulation of the angular velocity.

\bibitem{TBA02}K.L. Thompson, K.M.S. Bajaj, and G. Ahlers, Phys. Rev. E {\bf 65}, 046218 (2002).

\bibitem{ACBS94} G. Ahlers, D.S. Cannell, L. Berge, and S. Sakurai, Phys. Rev. E {\bf 49}, 545 (1994).

\bibitem{MA02} W. Meevasana and G. Ahlers, Phys. Rev. E {\bf 66}, 046308  (2002).

\bibitem{BBMTHCA96} J. Bruyn E. Bodenschatz, S. Morris, S. Trainoff, Y. Hu, D. Cannell, and G. Ahlers. Rev. Sci. Instrum. {\bf 67}, 2043 (1996).

\bibitem{TC03}S.P.  Trainoff and D.S. Cannell, Phys. Fluids {\bf 14}, 1340 (2003).


\end{thebibliography}
\end{document}